\documentstyle[a4p,epsfig,12pt]{article}

\newcommand{\sqee}{\sqrt{s_{\rm ee}}}

\def\ee{\mbox{e}^+\mbox{e}^-}
\def\sqee{\sqrt{s_{\rm ee}}}

\def\q{\rm q}
\def\g{\rm g}
\def\qmax{Q^2_{\rm max}}

\def\qqbar{\mbox{q}\overline{\mbox{q}}}

\def\pz{\phantom{0}}

\def\gg{\gamma\gamma}
\def\see{\sqrt{s_{{\rm ee}}}}
\def\rphi{r\phi}
\def\wvis{W_{\rm vis}}
\def\ptg{p_{\rm T}^\gamma}
\def\etag{\eta^\gamma}
\def\ptjet{p_{\rm T}^{\rm jet}}
\def\etajet{\eta^{\rm jet}}
\def\piz{\pi^0}
\def\fmax{f_{\rm max}}
\def\scl{\sigma_{\rm cluster}}
\def\nbar{\bar{\rm n}}

\def\lumi{\mathcal{L}}
\def\xtg{x_{\rm T}^\gamma}
\def\NEV{137}
\def\FGAMMA{a=0.86 \pm 0.08{\rm ~(stat)}}
\def\FPI0{b=0.12 \pm 0.08 {\rm ~(stat)}}

\def\XSECTION{\sigma_{{\rm tot}}=0.32 \pm 0.04 {\rm ~(stat)} \pm 0.04 {\rm ~(sys)} \rm ~pb}

\def\EPSSR{(51.8\pm0.7~(\rm stat))\%}
\def\EPSDR{(61.8\pm1.1~(\rm stat))\%}

\def\NPROMPT{110.8}

\def\rsinglexll{r= 0.47 \pm 0.11 {\rm ~(stat)}} 
\def\rsinglexg{r= 0.64 \pm 0.12 {\rm ~(stat)}}
\def\rsinglextg{r= 0.43 \pm 0.08 {\rm ~(stat)}}
\def\SYSETANBAR{2.0\%}
\def\SYSREW{2.8\%}
\def\SYSBIN{1.3\%}
\def\SYSMINBL{1.8\%}
\def\SYSMCSTAT{1.6\%}
\def\SYSECAL{7.0\%}
\def\SYSPI0{5.5\%}
\def\SYSPI{8.3\%}

\def\SYSANG{0.2\%}
\def\SYSC{0.5 \%}
\def\SYSSTD{4.0\%}
\def\SYSHERWIG{3.8\%}
\def\SYSPDF{3.8\%}
\def\dsdpt{{\rm d}\sigma/{\rm d}\ptg}
\def\dsdeta{{\rm d}\sigma/{\rm d}|\etag|}
\def\dsdxll{{\rm d}\sigma/{\rm d} x_{\rm LL}^-}
\def\dsdxtg{{\rm d}\sigma/{\rm d} x_{\rm T}^\gamma}
\begin{document}

\begin{titlepage}
\begin{center}{\large   EUROPEAN ORGANISATION FOR NUCLEAR RESEARCH
}\end{center}\bigskip
\begin{flushright}
 CERN-EP/2003-023   \\ 5th May 2003
\end{flushright}
\bigskip\bigskip\bigskip\bigskip\bigskip
\begin{center}{\huge\bf \boldmath  
Measurement of Isolated Prompt Photon Production in Photon-Photon Collisions
at $\sqee=183-209$~GeV 
}\end{center}\bigskip\bigskip
\begin{center}{\LARGE The OPAL Collaboration
}\end{center}\bigskip\bigskip
\bigskip\begin{center}{\large  Abstract}\end{center}
For the first time at LEP the production of prompt photons is studied in
the collisions of quasi-real photons using 
the OPAL data taken at $\ee$ centre-of-mass energies between 
183~GeV and 209~GeV. 
The total inclusive production cross-section for isolated prompt photons
in the kinematic range of transverse momentum 
$\ptg > 3.0$~GeV and pseudorapidity $|\etag|<1$
is determined to be $\XSECTION$.
Differential cross-sections are compared 
to the predictions of a next-to-leading-order (NLO) calculation.

\bigskip\bigskip\bigskip\bigskip
\bigskip\bigskip
\bigskip\bigskip\bigskip
\begin{center} {\large (To be submitted to Eur.~Phys.~J.~C)}
\end{center}

\end{titlepage}
\begin{center}{\Large        The OPAL Collaboration
}\end{center}\bigskip
\begin{center}{
G.\thinspace Abbiendi$^{  2}$,
C.\thinspace Ainsley$^{  5}$,
P.F.\thinspace {\AA}kesson$^{  3}$,
G.\thinspace Alexander$^{ 22}$,
J.\thinspace Allison$^{ 16}$,
P.\thinspace Amaral$^{  9}$, 
G.\thinspace Anagnostou$^{  1}$,
K.J.\thinspace Anderson$^{  9}$,
S.\thinspace Arcelli$^{  2}$,
S.\thinspace Asai$^{ 23}$
% ...
}\end{center}\bigskip
\bigskip
$^{  1}$School of Physics and Astronomy, University of Birmingham,
Birmingham B15 2TT, UK
\newline
$^{  2}$Dipartimento di Fisica dell' Universit\`a di Bologna and INFN,
I-40126 Bologna, Italy
\newline
$^{  3}$Physikalisches Institut, Universit\"at Bonn,
D-53115 Bonn, Germany
\newline
% ...
$^{  5}$Cavendish Laboratory, Cambridge CB3 0HE, UK
\newline
% ...
$^{  9}$Enrico Fermi Institute and Department of Physics,
University of Chicago, Chicago IL 60637, USA
\newline
% ...
$^{ 16}$Department of Physics, Schuster Laboratory, The University,
Manchester M13 9PL, UK
\newline
% ...
$^{ 22}$Department of Physics and Astronomy, Tel Aviv University,
Tel Aviv 69978, Israel
\newline
$^{ 23}$International Centre for Elementary Particle Physics and
Department of Physics, University of Tokyo, Tokyo 113-0033, and
Kobe University, Kobe 657-8501, Japan
\newline
% ...
\section{Introduction}
We present the first LEP measurement of the inclusive 
production of isolated prompt photons in photon-photon collisions, 
$\gg\to\gamma+\mbox{X}$, where X denotes the hadronic final
state produced in addition to the photon. 
The interacting photons are emitted by the beam 
electrons\footnote{Positrons are also referred to as electrons.}.
Electrons scattered at small angles into the beam pipe
escape detection and, in this analysis,
events with one or two detected scattered electrons are vetoed 
(``anti-tagging''). The interacting photons thus 
carry a small four-momentum squared, $Q^2$, i.e. they are quasi-real.

\begin{figure}[htbp]
   \begin{center}
      \mbox{
          \epsfxsize=6.5cm
          \epsffile{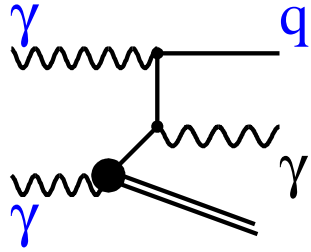}
           }
   \end{center}
\caption{
Diagram of the single-resolved process. 
The double line indicates the photon remnant and
the dark circle a resolved photon.
}
\label{fig-feyn1}
\end{figure}

\begin{figure}[htbp]
   \begin{center}
    \vskip -4mm
    \begin{tabular}{ll}
      \mbox{
          \epsfxsize=6.5cm
          \epsffile{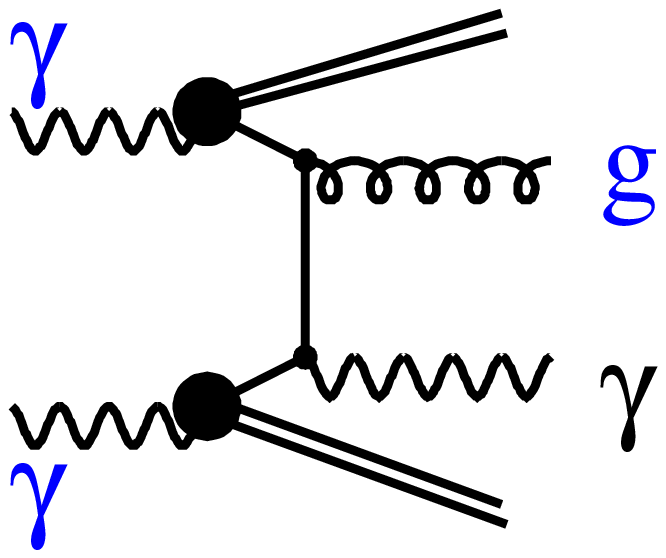}
           }
&
	\mbox{	
	  \epsfxsize=6.5cm	
          \epsffile{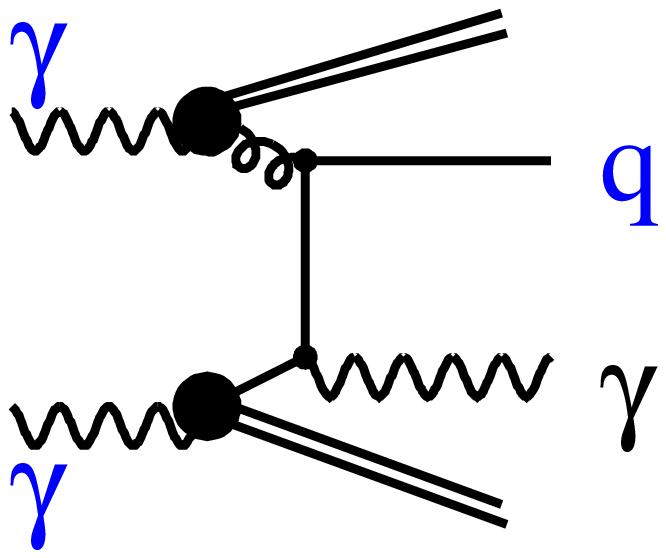}
	}
    \end{tabular}	
   \end{center}
\caption{
Diagrams of double-resolved processes.
The double lines indicate the photon remnant and
the dark circles a resolved photon.
}
\label{fig-feyn2}
\end{figure}

\begin{figure}[hbtp]
   \begin{center}
    \vskip -5mm
      \mbox{
          \epsfxsize=7.0cm
          \epsffile{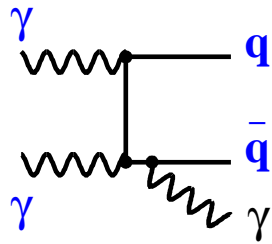}
           }
   \end{center}
\caption{Diagram of a direct process with Final State Radiation (FSR).}
\label{fig-feyn3}
\end{figure}

In leading order (LO), only processes where one (``single-resolved'') or 
both (``double-resolved'') of the incoming photons fluctuate into a 
hadronic state contribute to the production cross-section for prompt photons.
In these processes, a quark or a gluon from the hadronic state 
participates in the hard interaction, 
$\gamma \q\to\gamma \q$~(Fig.~\ref{fig-feyn1}), 
$\qqbar \to \gamma \g$ and $\g\q\to\gamma \q$~(Fig.~\ref{fig-feyn2}). 
Processes with final state radiation (``FSR'') 
are a higher-order contribution to the direct process (Fig.~\ref{fig-feyn3}).

The hadronic structure of the photon has previously been studied by OPAL in
the interactions of two quasi-real photons producing jets~\cite{bib-opaljet},
hadrons~\cite{bib-hadrons} or D$^{*}$ mesons~\cite{bib-dstar}
at high transverse momentum. The inclusive 
production cross-section for isolated prompt photons is expected to be 
about two orders of magnitude smaller than for di-jet production
in a similar kinematic region of transverse energy $E_{\rm T}>3$~GeV and 
pseudorapidity $|\eta|<1$~\cite{bib-opaljet}. 
Hadronisation uncertainties, however, are expected to be much smaller than 
in the case of jet production at similar transverse momenta,
allowing a complementary study of the hadronic
structure of photon interactions~\cite{bib-drees,bib-gordon}.
The photoproduction of prompt photons has previously 
been studied by NA14~\cite{bib-na14} and by ZEUS~\cite{bib-zeus}. 

Approximately $649 \rm ~pb^{-1}$ of $\ee$ collision data taken 
by the OPAL experiment at centre-of-mass energies $\sqee$
from 183 GeV to 209 GeV are used in this analysis.
Since the expected production
cross-section is small and the increase of the cross-section from 
the lowest to the highest centre-of-mass energy is expected to be
less than the statistical uncertainty of the measurement~\cite{bib-drees}, 
all data are combined for the final result. 
The luminosity-weighted mean $\ee$ centre-of-mass energy is 
approximately 196.6 GeV.
The measured differential and total cross-sections are compared to
the leading order predictions of the Monte Carlo generator 
PYTHIA~\cite{bib-pythia} and to
a next-to-leading-order (NLO) calculation~\cite{bib-fgh}.
The measurement is restricted to 
isolated prompt photons to suppress background from neutral particle decays
into photons. 

\section{The OPAL detector}

A detailed description of the OPAL detector
can be found in~\cite{bib-opaltechnicpapers}, and
therefore only a brief account of the main features relevant
to the present analysis will be given here.

The central tracking system is located inside a solenoidal magnet which
provides a uniform axial magnetic field of 0.435~T.
The magnet is surrounded by a lead-glass electromagnetic
calorimeter (ECAL) and a hadronic sampling calorimeter (HCAL).
The HCAL is surrounded by muon chambers.
There are similar layers of detectors in the
endcaps. The region around the beam pipe on both sides
of the detector is covered by the forward calorimeters and the
silicon-tungsten luminometers.

Starting with the innermost components, the
tracking system consists of a high precision silicon
microvertex detector (SI), a precision vertex
drift chamber (CV), a large volume jet chamber (CJ) with 159 layers of axial
anode wires and a set of $z$ chambers measuring the track coordinates
along the beam direction\footnote{
  In the OPAL coordinate system the $x$ axis points
  towards the centre of the LEP ring, the $y$ axis points upwards and
  the $z$ axis points in the direction of the electron beam.
  The polar angle $\theta$ is defined with respect to the $z$ axis.
  The azimuthal angle $\phi$
  and the radius $r$ denote the usual spherical coordinates.}.

The ECAL covers the complete azimuthal range for polar angles
satisfying $|\cos\theta|<0.98$. The barrel section, which covers
the polar angle range $|\cos\theta|<0.82$, consists of a cylindrical array
of 9440 lead-glass blocks with a depth of
$24.6$ radiation lengths. Each block subtends an angular region of 
approximately $40 \times 40$~mrad$^{2}$. Deposits of energy in
adjacent blocks are grouped together to form clusters. The intrinsic
energy resolution of $\sigma_{E}/E=0.2\% \oplus 6.3\% / \sqrt{E}$
is substantially degraded (by a factor $\simeq 2$) due to
the presence of at least two radiation lengths of material in front of the
lead glass.

The endcap sections consist of 1132 
lead-glass blocks with a depth of more than $22$ radiation lengths, 
covering the range of polar angles between $0.81 < |\cos\theta| < 0.98 $.
The HCAL consists of streamer tubes and thin multiwire chambers 
instrumenting the gaps in the iron yoke of the magnet, which provides
the absorber material of 4 or more interaction lengths.

Scintillators in the barrel and endcap regions 
provide time measurements for the large fraction of
photons which convert in the material in front of the ECAL.
They are also used to reject the background from cosmic ray 
interactions. 
The barrel time-of-flight (TOF) scintillator bars
are located outside the solenoid 
in front of the barrel ECAL
and match  
its geometrical 
acceptance ($|\cos{\theta}| < 0.82$).

The forward calorimeters (FD) at each end of the OPAL detector
consist of cylindrical lead-scintillator calorimeters with a depth of
24 radiation lengths.
The acceptance of the forward calorimeters covers the angular range
from 47 to 140~mrad from the beam direction.
The silicon tungsten detectors (SW)~\cite{bib-opalsiw} at each end of the OPAL
detector cover the angular region between 25 mrad and 59 mrad
in front of the forward calorimeters.
Due to a radiation shield installed for LEP2 running, the lower edge of
the useful acceptance is 33 mrad.
Each calorimeter consists of 19 layers of silicon detectors and 18
layers of tungsten, corresponding to a total of 22 radiation
lengths. 

\section{Process kinematics and Monte Carlo simulation} 

The kinematic properties of the two interacting photons are described by their
negative squared four-mo\-mentum transfers, $Q^2_{i=1,2}$, which
are related to the scattering angles $\theta'_i$  of the
corresponding electron relative to the beam direction by
\begin{equation}
Q^2_i  =-(k_i-k'_i)^2\approx 2E_i E'_i(1-\cos\theta'_i),
\label{eq-q2}
\end{equation}
neglecting the mass $m_{\rm e}$ of the electron.
The quantities $k_i$ and $k'_i$ are the four-momenta of the beam
and scattered electrons, and $E_i$ and $E'_i$ their energies. 
The maximum squared four-momentum transfer, $\qmax$, is  
given by Eq.~\ref{eq-q2} using the limits on the energy and
the polar angle from the anti-tagging requirement that no electron has been
detected in the ECAL, FD or SW calorimeters.
These cuts correspond to $\qmax \approx 10$~GeV$^2$. 
No correction for this anti-tagging condition is applied.
The median $Q^2$ resulting from this definition cannot be determined with the
data since the scattered electrons are not tagged. For the kinematic
range of this analysis the Monte Carlo simulations predict the median $Q^2$
to be of the order $10^{-4}$~GeV$^2$.

The Monte Carlo generator PYTHIA~6.130~\cite{bib-pythia} is used for the 
simulation of signal events, $\gg \to \gamma + \mbox{X}$.
Separate samples of single and double-resolved processes have been 
produced using the SaS-1D~\cite{bib-sas}, GRV-LO~\cite{bib-grv}, and 
LAC-1~\cite{bib-lac} parametrisations 
of the parton distributions of the photon.
The samples are generated at $\ee$ centre-of-mass energies 
$\see$ of 189~GeV, 192~GeV, and 206~GeV.

Several samples are generated for systematic studies:
single-resolved signal events with HERWIG~\cite{bib-HERWIG}
and signal events with final state photon radiation with 
PHOJET~1.10~\cite{bib-phojet} using
the parton shower option of JETSET~\cite{bib-pythia}.
Photon-photon events with initial state photon radiation
are simulated using BDK~\cite{bib-bdk}.

The transition from quasi-real to virtual photons is smooth and
the definition used to separate these kinematic regions is somewhat arbitrary.
In the Monte Carlo simulation we choose values of $Q^2<4.5$~GeV$^2$ to define 
quasi-real photons. 
The background from hadronic $\gg$ events without prompt photon production is
simulated with PHOJET for the case where both photons have $Q^2<4.5$~GeV$^2$.
In PHOJET the fragmentation into hadrons is performed using JETSET. 
HERWIG~5.9 is used to simulate $\gg$ events where one photon has 
$Q^2<4.5$~GeV$^2$ and the other photon has $Q^2>4.5$~GeV$^2$.
This combination of generators gives a good description
of hadronic two-photon events~\cite{bib-opaljet}.
Two-photon events with fully leptonic final states are simulated
with Vermaseren 1.0~\cite{bib-Vermaseren}. 

PYTHIA 5.7 and KK2f~\cite{bib-kk2f} are used for the 
description of $\qqbar(\gamma)$ events produced in $\ee$ annihilations. 
KORALZ 4.02~\cite{bib-koralz} and KK2f are used to simulate $\tau$ pairs.
Four-fermion final states are simulated with grc4f~2.1~\cite{bib-GRC4F} and 
KORALW~\cite{bib-koralw}. 

A photon signal in the electromagnetic calorimeter can   
originate from a prompt photon but also from 
$\piz$ or $\eta$ mesons decaying into two photons.
Background is also expected from the annihilation of antineutrons, $\nbar$,
in the detector material.
The different distributions of energy deposited in the calorimeter for signal
and background can be used to separate signal from background
by a cluster shape analysis.
To study the response of the electromagnetic calorimeter to 
the various sources of background,
Monte Carlo samples are generated using a single-particle generator.
These particles are generated with a flat $p_{\rm T}$ distribution 
in the range of 2 GeV to 13 GeV, and flat azimuthal angle $\phi$ and
flat pseudorapidity $\eta = -\ln( \tan (\theta / 2))$ distributions over 
the acceptance of the barrel part of the detector.
Exponential $p_{\rm T }$-dependent weighting functions are introduced to
reproduce the shape of the $p_{\rm T }$ distributions as predicted by PHOJET. 

All Monte Carlo samples are passed through a full simulation of the 
OPAL detector~\cite{bib-gopal} and
are analysed using the same reconstruction algorithms as
for data.

\section{Event selection}
Only anti-tagged photon-photon scattering events are studied in this analysis. 
The same event selection is applied to the data samples taken
at different e$^+$e$^-$ centre-of-mass energies $\sqee$.
Anti-tagged photon-photon events are selected using the following 
requirements:
\begin{itemize}
\item Anti-tagged events are selected by vetoing all events containing
      an energy deposit of more than $15\%$ of $\sqee$ in the SW calorimeter
      or more than $25\%$ of $\sqee$ in the FD calorimeter, 
      in either hemisphere of the detector.
      This corresponds to a maximum allowed scattering angle of
      the beam electrons of $\theta'=33~{\rm mrad}$ for electrons
      with energies above the threshold.
\item At least three tracks must be found in the tracking chambers. 
      A track is
      required to have a minimum transverse momentum of 120~MeV and 
      more than 20 hits in the central jet chamber.
      In this paper, transverse
      is always defined with respect to the $z$ direction of the detector.
      The distance of closest approach to 
      the origin must be less than 20~cm in the $z$ direction and less than 
      2.5~cm in the $r\phi$ plane.
      An event is rejected if a track with a momentum higher than $25 \%$ 
      of the centre-of-mass energy is detected.
\item 
      A minimum visible invariant mass of the event,
      $W_{\rm vis}$, of more than $5$ GeV is required.
      To reduce the $\ee$ annihilation background,
      $W_{\rm vis}$ should be less than $30\%$ of $\sqee$. 
      $W_{\rm vis}$ is calculated using the
      energies and positions of clusters measured in the ECAL, HCAL, 
      FD and SW calorimeters and using 
      the momenta of tracks. A matching algorithm~\cite{bib-matching}
      is applied to compensate for double-counting of particle momenta in
      the calorimeters and in the tracking chambers.
\item The background due to beam-gas or beam-wall interactions is
      reduced by requiring the absolute value of the net charge of an event, 
      calculated by adding the charges of the tracks, to be less than
      three. In addition, the ratio of the longitudinal component
      of the momentum vector of the final state to
      the total visible energy is required to be smaller than $0.98$. 
\end{itemize}  
Photon candidates are selected as follows:
\begin{itemize}
 \item 
   The photon candidates are reconstructed using the same algorithm as 
   in~\cite{bib-cg}. Only clusters in the ECAL which consist
   of at least $2$ lead-glass blocks and fewer than $13$ blocks are taken.  
   If a track is associated with the cluster, the candidate is rejected.
   The cluster has to be totally contained in 
   the barrel part of the calorimeter; clusters with blocks in the end cap
   calorimeter are rejected. 
   The pseudorapidity of the photon, $\etag$, is required to be in 
   the range $-1 < \etag< 1$. 
   The minimum transverse momentum of the photon candidate with respect to
   the beam axis, $\ptg$, is $3$ GeV. 
   The energy $E_{\gamma}$ and the polar angle $\theta_{\gamma}$
   of the photon candidate are calculated from
   the energy and position of the cluster.
 \item 
   If hits in the time-of-flight (TOF) detector are associated with 
   the cluster,
   the measured time at the TOF is required to be less than
   $2$~ns from the expected arrival time of
   a photon originating from the beam crossing. 
   This cut rejects cosmic ray
   events and also removes background from antineutrons
   produced in photon-photon interactions.
 \item 
   We apply the isolation criterion proposed in~\cite{bib-frixione}
   to the photon candidate using all detected particles based
   on the matching algorithm~\cite{bib-matching}.   
   For each particle $i$, the distance
   \begin{equation}
 R_{i\gamma}=\sqrt{(\phi_i-\phi_\gamma)^2 + (\eta_i-\eta_\gamma)^2}
   \end{equation}
   to the photon candidate is computed in $\phi\eta$ space, 
   where $\phi$ and $\eta$ are the azimuthal angle
   and pseudorapidity, respectively.
   A photon candidate is kept if the condition
   \begin{equation}
     \sum_{{\mathrm particles},i}{E_{{\rm T},i}\Theta(\delta - R_{i,\gamma})} \le 
              0.2 \cdot E_{{\rm T},\gamma} \frac{1-\cos(\delta)}{1-\cos(R)},\mbox{ for all }
      \delta \le R 
   \label{eq-frix}
   \end{equation}
    is fulfilled,
    where $E_{{\rm T},i}$ is the transverse energy of the $i{^{th}}$ particle,
     $\Theta$ is the step function, which ensures that only particles in
     the cone with opening half-angle $\delta$ contribute to the sum, and
     the cone radius $R=1$.  
   Events with more than one isolated photon are rejected.
 \item 
   The difference between the azimuthal angle $\phi^{\gamma}$ 
   of the photon candidate 
   and the azimuthal angle of the remaining hadronic
   system is required to be between $\pi-1$ and $\pi+1$.
 \item 
   Background from events with purely leptonic final states is
   further reduced by rejecting events where more than $50\%$ of all tracks 
   are identified as electrons using the specific
   energy loss, ${\mathrm d}E/{\mathrm d}x$, in the jet chamber.
   For this identification,
   the ${\mathrm d}E/{\mathrm d}x$ probability 
   for the electron hypothesis should exceed $50\%$.  
\end{itemize}
After applying all cuts $\NEV$ data events are selected.

\section{Determination of the number of photons}
\begin{figure}[htbp]
   \begin{center}
      \mbox{
          \epsfxsize=15.0cm
          \epsffile{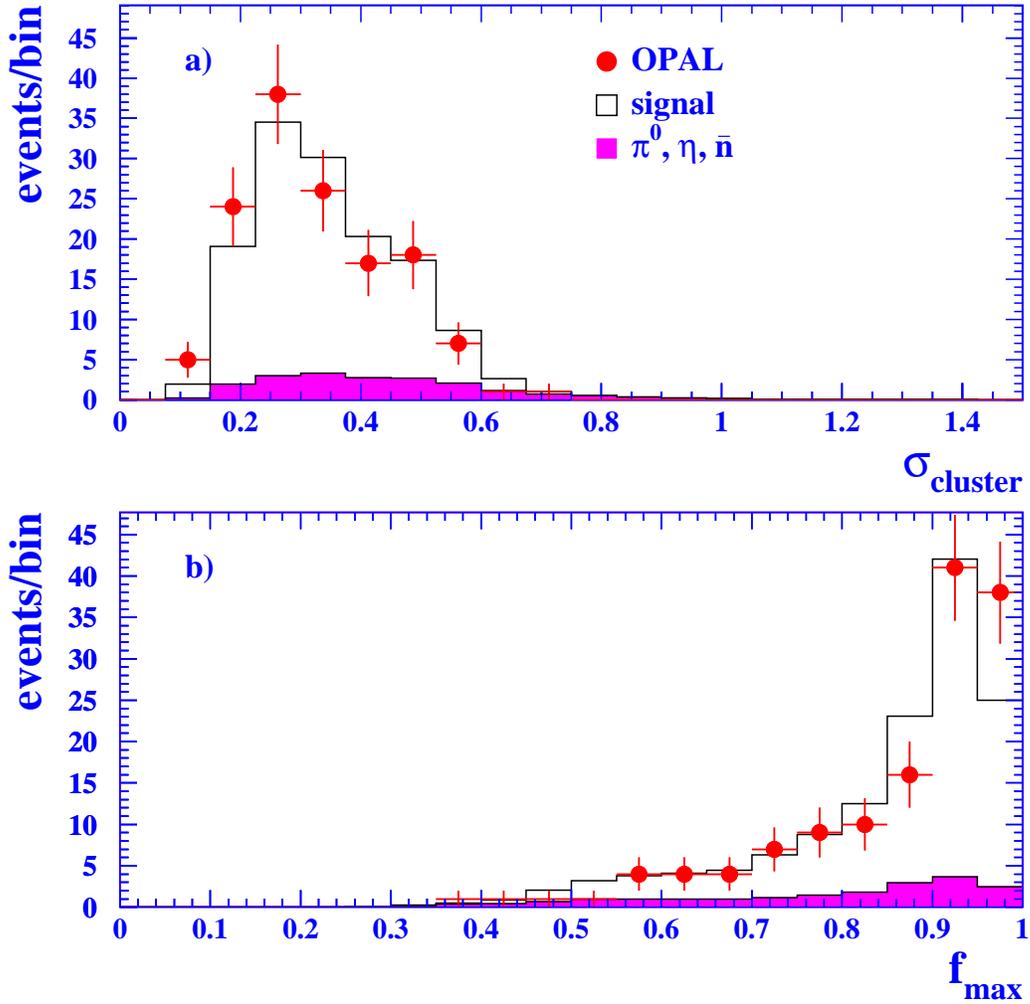}
           }
   \end{center}
\caption{
Distributions of the 
shower-shape variables for data and Monte Carlo events: 
a) sum of the energy-weighted quadratic deviations of the 
lead-glass block coordinates with respect to the coordinates of the cluster,
$\scl$;
b) ratio of the energy of the 
most energetic block of the cluster to the total cluster energy, $\fmax$.
The ratio of signal to background is determined by a binned maximum
likelihood fit to the two-dimensional distribution.
}
\label{fig-shape}
\end{figure}

The main background to the prompt photon signal is from photons 
produced in $\piz$ and $\eta$ decays, and from antineutrons, $\nbar$.
To separate signal photons from the background, a cluster shape 
analysis is performed. The cluster shape analysis derives
the background rate from the data and is independent
of the Monte Carlo predictions for the background rates.
Two cluster shape variables are used:
\begin{itemize}
 \item The sum of the energy-weighted quadratic deviations of the 
   lead-glass block
   coordinates with respect to the coordinates of the cluster,
   \begin{equation}
      \scl = 
       \frac {\sum_{{\mathrm blocks,}i}E_i((\phi_i-\phi_{\gamma})^2+
         (\theta_i-\theta_{\gamma})^2)} {E_{{\gamma}}}.
   \end{equation} 
 \item The ratio $\fmax$ of the energy of the most energetic block
   of the cluster to the total cluster energy,
   \begin{equation}
        f_{{\rm max}}=\frac{E_{{\rm max}}}{E_{{\gamma}}}.
   \end{equation} 
\end{itemize}
To obtain the fraction of prompt photons in the sample
of candidates, the normalised two-dimensional distribution 
of $\fmax$ and $\scl$
is parametrised as a sum of signal and background 
contributions:
\begin{eqnarray}
\nonumber\lefteqn{g(f_{\rm max},\scl)=}\\
\nonumber & & 
a g^\gamma(f_{\rm max},\scl) +
                            b g^{\pi^0}(f_{\rm max},\scl) + \\
& & 
                  (1-a-b) ( c g^{\eta}(f_{\rm max},\scl) +
                            (1-c) g^{{\rm \bar{n}}}(f_{\rm max},\scl)),
\label{eq-fit}
\end{eqnarray}
where  $g^{k}(f_{\rm max},\scl)$ denotes the normalised
distribution obtained for particle type $k$; $a$ and $b$ denote the
$\gamma$ and $\pi^0$ fractions in the candidate sample
while $c$ is fixed to match the ratio of events, 
$N(\eta)/(N(\bar{{\rm n}})+N(\eta))$, 
predicted by PHOJET for  
photon-photon events with the same selection criteria as are applied 
to the data.
The one-dimensional distributions of $\fmax$ and $\scl$ are shown
in Fig.~\ref{fig-shape}. 
The shape of the signal distribution is taken from the detector simulation
of the prompt photon signal using PYTHIA.  
The simulation of the shower-shape variables  
for single photons has been compared to the shower-shape variables
measured for photons in radiative Bhabha events, 
$\ee\to\ee\gamma$, selected from the OPAL data, and is
found to be consistent.

A binned maximum likelihood fit is applied to determine the fractions
$a$ and $b$, 
assuming that the content of each bin follows a Poisson distribution.
The fit yields a photon contribution of $\FGAMMA$ and a $\piz$ 
contribution of $\FPI0$, where the uncertainties are due to
the statistical uncertainties of the data.
Within this uncertainty the $\pi^0$ rate
is consistent with the $\pi^0$ production cross-section predicted by PHOJET. 

\section{Separation of single and double-resolved events}
\label{sec-sep}
To study the relative contributions of single and double-resolved processes
the variables 
\begin{eqnarray}
x_{\gamma}^{\pm} 
&=& \frac{p_{\rm T}^{\gamma} e^{\pm\etag } + p_{\rm T}^{\rm jet} e^{\pm\eta^{\rm jet} }}{y^\pm\see} 
\approx 
\frac{p_{\rm T}^{\gamma} e^{\pm\etag } + p_{\rm T}^{\rm jet} e^{\pm\eta^{\rm jet} }}
{\Sigma_{\rm hadrons,\gamma}(E\pm p_z)}. 
\label{eq-xpm}
\end{eqnarray}
can be defined in $\gamma$ plus jet events, where
$y^{\pm}=E_{\gamma}/E_{\rm e}$ 
are the fractional energies of the quasi-real initial photons
oriented towards the positive and negative $z$ axis, and
$p_{\rm T}$, $\eta$ are the transverse momenta and pseudorapidities
of the jet and prompt photon, respectively.
Since $y^\pm$ cannot be measured directly, the denominator of Eq.~\ref{eq-xpm}
is approximated by summing over
the energies $E$ and the $z$ components of the momenta, $p_{z}$,
of all detected final state particles.

The variables $x_{\gamma}^{+}$ and $x_{\gamma}^{-}$ are measures of the 
fractions of the initial photons momenta participating in the hard 
interaction. 
In LO, $x_{\gamma}^+$ and $x_{\gamma}^-$ should be 
smaller than~1 for double-resolved events, 
whereas for single-resolved events only one of two variables
is smaller than~1 and the other variable, related to the
directly interacting photon, equals~1.

In this paper we use similar variables proposed in~\cite{bib-fgh}
\begin{eqnarray}
x_{\rm LL}^{\pm} 
&=&\frac{p_{\rm T}^{\gamma} (e^{\pm \etag } +e^{\pm \eta^{{\rm jet}} })}
{y^\pm \see}
\approx \frac{p_{\rm T}^{\gamma} (e^{\pm \etag } +e^{\pm \eta^{{\rm jet}} })}
{\Sigma_{\rm hadrons,\gamma}(E\pm p_z)}. 
\label{eq-xll}
\end{eqnarray}
where the transverse momentum of the
jet has been replaced by the transverse momentum of the prompt photon.
Since the transverse momentum is measured with better
resolution for a photon than for a jet, 
the experimental resolution for $x_{\rm LL}^{\pm}$ is better than for
$x_{\gamma}^{\pm}$.

The jets are reconstructed using a cone algorithm~\cite{bib-coneopal}.
They are required to have a transverse momentum of $\ptjet>2.5$~GeV,
the pseudorapidity $\etajet$ of the jet must be in the range 
$\left[-2,2\right]$, and 
the radius of the cone in $\eta\phi$ space is set to $R=1$.
A lower cut for $\ptjet$ than for $\ptg$ is chosen because
a symmetric cut $\ptjet=\ptg$ leads to infrared 
instabilities in the NLO calculations.

About $64\%$ of the selected events with a photon have exactly one jet
and about $20\%$ have two or more jets. 
Events with two or more jets are included in the $\gamma$ plus 
jet sample by evaluating $x_{\rm LL}^{\pm}$ 
using the jet with the highest transverse momentum.

The fraction of single-resolved events in the $\gamma$ plus jet
sample is determined by a binned maximum likelihood fit to 
the normalised two-dimensional $x_{\rm LL}^{\pm}$ distribution.
\begin{equation}
 g(x_{\rm LL}^+,x_{\rm LL}^-)=r a g^{\rm sr}(x_{\rm LL}^+,x_{\rm LL}^-) 
+(1-r)a 
g^{\rm dr}(x_{\rm LL}^+,x_{\rm LL}^-) + (1-a) g^{\rm bg}(x_{\rm LL}^+,x_{\rm LL}^-).
\end{equation}
In the fit the sum of the Monte Carlo distributions 
for single (sr) and double-resolved (dr) events is fixed to the
number of prompt photons in the data, obtained by the shower-shape analysis.
The fraction $1-a$ of $\pi^0$, $\eta$, and $\nbar$
background (bg) events is also fixed to the value derived from the
shower-shape fit. The shape of the $x_{\rm LL}^{\pm}$ distributions
for the single and double-resolved events
is taken from PYTHIA and for the background events from PHOJET.
The only free parameter is the fraction $r$ of single-resolved events
in the selected data sample.
The fit yields a fraction of single-resolved events $\rsinglexll$.

\begin{figure}[htbp]
   \begin{center}
      \mbox{
          \epsfxsize=15.0cm
          \epsffile{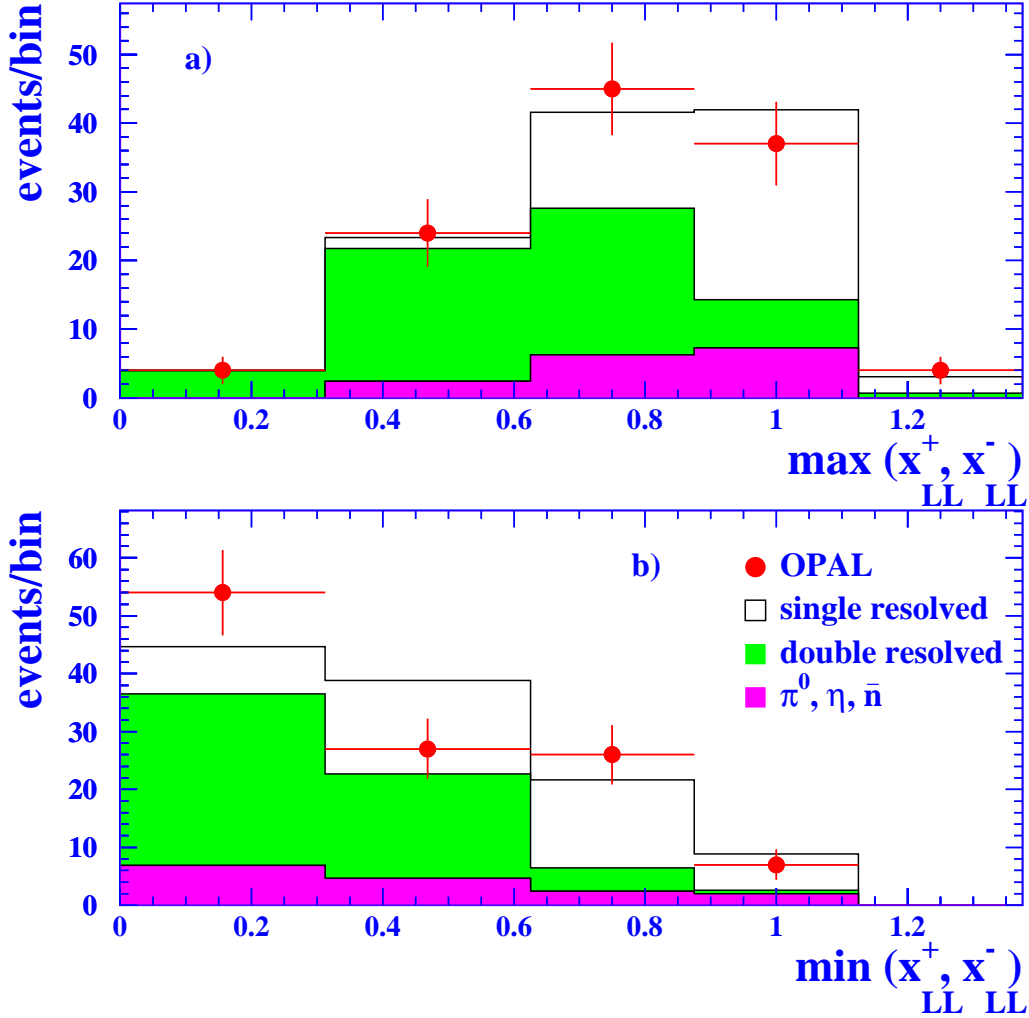}
           }
   \end{center}
\caption{ 
 Distribution of the a) maximum and b) minimum of $x_{\rm LL}^+$ and 
 $x_{\rm LL}^-$ for the selected $\gamma$ plus jet events. 
 The uncorrected data distributions are compared to the sum of the
 single and double-resolved signal Monte Carlo distributions, 
 and the background Monte Carlo distributions. 
 The sum of the signal and background Monte Carlo distributions is 
 normalized to the data.
 The background fraction is taken from the shower-shape fit and 
 the fraction of single-resolved signal events is taken from
 the fit to the $x_{\rm LL}^{\pm}$ distribution.
}
\label{fig-xg}
\end{figure}

In the original PYTHIA simulation the rate of single-resolved events
is predicted to be one order of magnitude larger than
the rate of double-resolved events. 
A good description of the data by the Monte Carlo simulation
is necessary to determine the correction factors 
for acceptance losses and resolution effects.
The rate of single-resolved events in PYTHIA after the detector
simulation and event selection is therefore
adjusted to the fitted value $\rsinglexll$.
Fig.~\ref{fig-xg} shows the 
$x_{\rm LL}^{\rm max}={\rm max}(x_{\rm LL}^+,x_{\rm LL}^-)$
and $x_{\rm LL}^{\rm min}={\rm min} (x_{\rm LL}^+,x_{\rm LL}^-)$ distributions
compared to the PYTHIA simulation using $\rsinglexll$.
The sum of the signal and background
Monte Carlo distributions is normalized to the data.
The fraction of the $\pi^0$, $\eta$ and $\nbar$ background 
is taken from the shower-shape fit. 

The $x_{\rm LL}^{\rm max}$ distribution is described well by the
sum of signal and background Monte Carlo after the fit, 
whereas the $x_{\rm LL}^{\rm min}$ 
distribution has a slight enhancement of low values of $x_{\rm LL}^{\rm min}$,
in the region where both single and double-resolved events contribute.

A variable similar to $x_{\rm LL}^{\pm}$ 
which is defined for all prompt photon events, not just
the subsample with jets, is 
the scaled transverse momentum $\xtg$ of the prompt photon. It is given by
\begin{equation}
\label{eq-xtg}
\xtg=\frac{2\ptg}{W}.
\end{equation}
For events with a photon and a centrally produced jet 
($\etag=\eta^{\rm jet}=0$), 
the variable $\xtg$ is equal to $x_{\gamma}^{\pm}$.
This variable is therefore also sensitive to the fractions of 
the single-resolved 
and the double-resolved processes. As with the $x_{\gamma}^{\pm}$
distribution, the single-resolved contribution dominates at higher values of 
$\xtg$, whereas
the double-resolved events are concentrated at smaller $\xtg$,
as predicted by the Monte Carlo. 

\section{Systematic uncertainties}
\begin{figure}[htbp]
   \begin{center}
      \mbox{
          \epsfxsize=15.0cm
          \epsffile{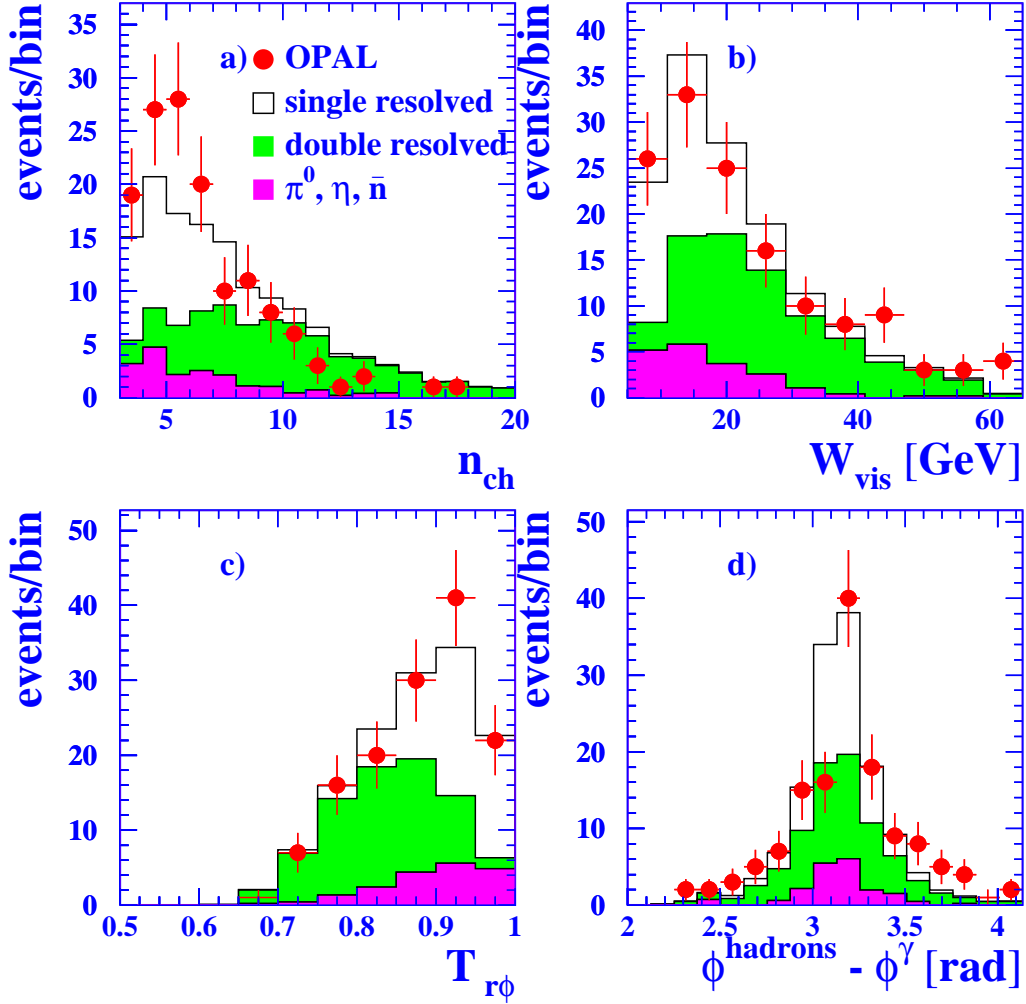}
           }
   \end{center}
\caption{
  Distributions of 
     a) the charged multiplicity, $n_{\rm ch}$,  
     b) the visible invariant mass,  $\wvis$, 
     c) the thrust in the $\rphi$ plane, $T_{r\phi}$,
     and d) the angle between the prompt photon and the remaining 
        hadronic system in the $\rphi$ plane.
 The sum of the signal and background Monte Carlo distributions is 
 normalized to the data.
 The background fraction is taken from the shower-shape fit and 
 the fraction of single-resolved signal events is taken from
 the fit to the $x_{\rm LL}^{\pm}$ distribution.
}
\label{fig-check}
\end{figure}

\begin{figure}[htbp]
   \begin{center}
      \mbox{
          \epsfxsize=15.0cm
          \epsffile{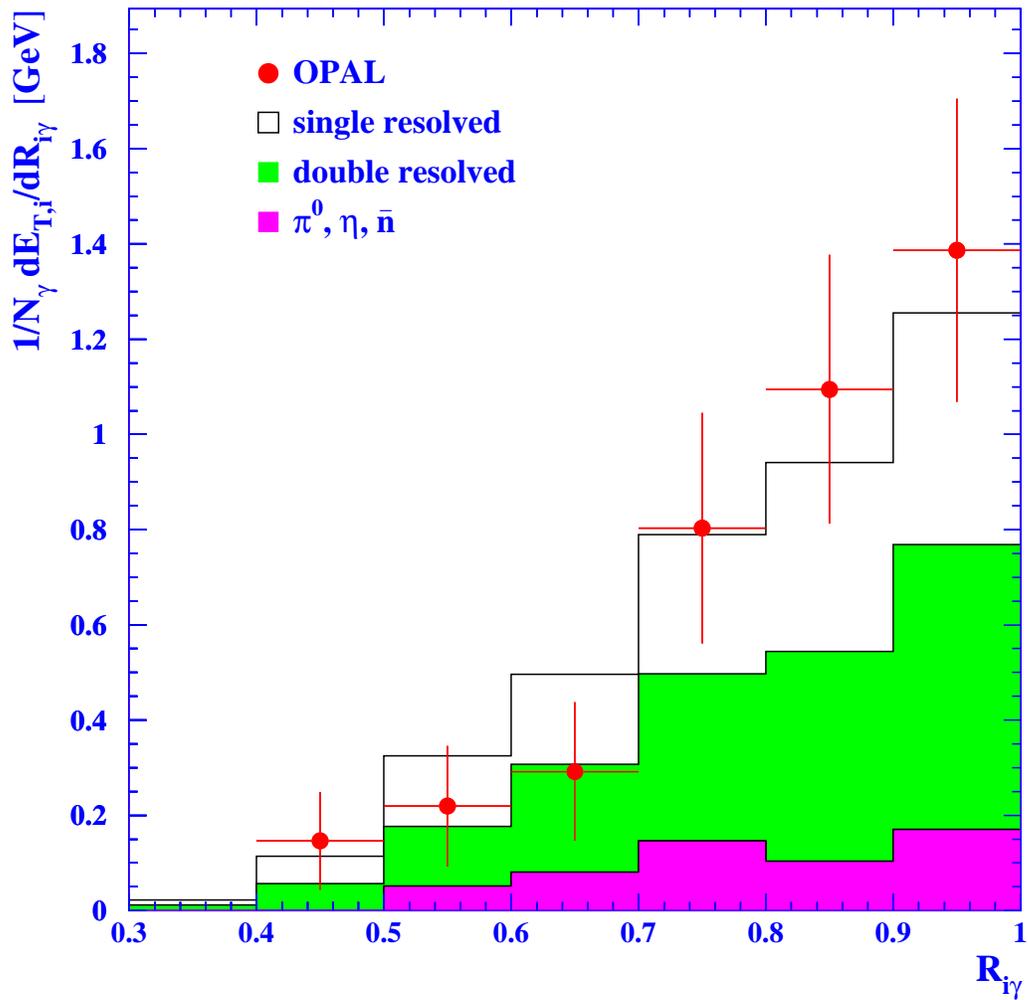}
           }
   \end{center}
\caption{ 
 Transverse energy flow around the isolation cone
 for the selected events.
 The background fraction is taken from the shower-shape fit and 
 the fraction of single-resolved signal events is taken from
 the fit to the $x_{\rm LL}^{\pm}$ distribution.
}
\label{fig-flow}
\end{figure}

Several kinematic distributions are shown in Fig.~\ref{fig-check}
to demonstrate the general agreement between the Monte Carlo and
data distributions.
The distributions of the 
charged multiplicity, $n_{\rm ch}$, the visible invariant mass,  $\wvis$, 
the thrust in the $\rphi$ plane, $T_{r\phi}$,
and the angle between the prompt photon and the remaining 
hadronic system in the $\rphi$ plane are shown after the event selection.
The sum of the signal and background Monte Carlo is normalised to the
data.
The transverse energy flow around the isolation cone,
$1/N_{\gamma} \cdot {\rm d} E_{T,i}/ {\rm d}R_{i\gamma}$,
is shown in Fig.~\ref{fig-flow}. 
The fraction of single-resolved events is taken 
from the fit to the $x_{\rm LL}^{\pm}$ distribution and
the fraction of $\pi^0$, $\eta$ and $\nbar$ background 
is taken from the shower-shape fit (Eq.~\ref{eq-fit}).
The shape of the $\pi^0$, $\eta$ and $\nbar$ background distributions
is simulated using PHOJET.
The Monte Carlo simulation describes the data well for all the distributions
shown.

The following systematic uncertainties are studied in detail:
\begin{itemize} 
\item 
 The background from $\pi^0$ decays with a single photon in 
 the isolation cone is irreducible. It is determined to be
 $N_{\piz \to 1\gamma}^{\rm data}=14 \pm 10$~(stat)
 using the fit to the shower-shape variables.
 The uncertainty on this background is estimated using the following procedure:
 \begin{itemize}
 \item 
  A $\piz$ peak is reconstructed using data events
  with two photons in the cone.
  Taking the ratio of $\piz$ events with one and two photons in the cone
  from the PHOJET simulation, the $\piz$ background is estimated to be
\begin{equation}
  N_{\piz \to 1\gamma}^{\rm data}=N_{\piz\to 2\gamma}^{\rm data}
  \frac{N_{\piz\to 1\gamma}^{\rm PHOJET}}{N_{\piz\to 2\gamma}^{\rm PHOJET}}
  =18 \pm 3~\mbox{(stat)}.
\end{equation}
   \item
   Due to isospin conservation the cross-sections
   and fragmentation functions for $\piz$ and $\pi^{\pm}$ production
   are expected to be proportional to each other.
   The analysis is therefore redone with the same selection cuts except that
   an isolated track has to be found instead of a photon. 
   The track is identified as a charged pion using
   the specific energy loss ${\rm d}E/{\rm d}x$ measured in the jet chamber.
   With these cuts the measured ratio of the number of charged to neutral 
   pions is found to be $11 \pm 3$~(stat) for the PHOJET simulation and 
   $12 \pm 9$~(stat) for the data.
 \end{itemize}
  An uncertainty of $50\%$ is assigned to the $\pi^0$ 
  background rate which is mainly due to the large statistical
  uncertainty of the tests performed.
  The resulting uncertainty on the total cross-section is $\SYSPI$.
\item
 The influence of the calibration of the ECAL on the selection efficiencies 
 is determined by varying the energy of 
 the lead-glass blocks by $\pm 3 \%$ for the data~\cite{bib-sigtot}. 
 The efficiencies obtained are compared to the original values
 and the difference is assigned as systematic uncertainty of $\SYSECAL$.
\item
 The fraction of single-resolved events is determined with the
 variable $x_{\rm LL}^{\pm}$ to be $\rsinglexll$.
 The systematic uncertainty is determined by fitting the $x_{\gamma}$ and
 $\xtg$ distributions defined in Section~\ref{sec-sep}.
 These fits yield $\rsinglexg$ and $\rsinglextg$, respectively.
 The value of $r$ is varied in the range $0.30<r<0.67$ within the 
 full uncertainty given by the largest deviation of the two fit results 
 from the value $r=0.47$ used in the analysis.
 Since the efficiencies are not very different for single and 
 double-resolved events, this
 leads to an uncertainty of only $\SYSSTD$ for the total cross-section. 
 \item
 The influence of the modeling of the parton density functions is studied
 by using Monte Carlo samples generated with the GRV-LO and LAC-1 
 parametrisations of the parton density functions. This yields a 
 systematic uncertainty of $\SYSPDF$.
 \item
 Using HERWIG instead of PYTHIA for the simulation of single-resolved events 
 changes the measured prompt photon production cross-section by $\SYSHERWIG$.
\item 
 To estimate the effect of the fixed $\eta$ to $\nbar$ ratio, 
 fits to the distributions of the shower-shape variables are 
 applied with either no $\nbar$ or no $\eta$ background. This affects the
 signal-to-background ratio by $\SYSETANBAR$.
\item 
 The sensitivity of the shower-shape variables to the exponential 
 weighting functions
 for the distributions of the single particle 
 generator events is estimated as follows. The parameters of the 
 exponential functions are determined by a fit of the single 
 particle generator $p_{\rm T}$ distribution
 to the PHOJET $p_{\rm T}$ distributions for the same particle type.
 The parameters of the weighting function
 are scaled to  $(1 \pm \alpha)$, where $\alpha$ is the relative error of 
 the fitted weighting parameter.
 The shower-shape distributions are re-weighted with the functions
 using the scaled parameter, and the analysis is redone. 
 The resulting systematic uncertainty is $\SYSREW$.
 \item The dependence of the shower-shape variables on the number of 
 required lead-glass blocks is studied by increasing the cut on the
 number of lead-glass blocks from 2 to 3 in the Monte Carlo but not
 in the data.
 This changes the signal-to-background ratio by $\SYSMINBL$.
\item
 The finite number of Monte Carlo events
 yields a systematic error of $\SYSMCSTAT$.
\item 
 The fits to the distributions of the shower-shape variables 
 are performed with various bin sizes and upper and lower bounds of
 the histograms. The bins sizes were doubled/halved and the histogram bounds
 were shifted by half a bin size.
 This changes the signal contribution parameter $a$ by $\SYSBIN$.
\item 
 The analysis is repeated using only the cluster shape 
 variable $C$~\cite{bib-C}. The variable
 $C$ is the result of a Monte Carlo fit which compares the measured 
 to the expected energies. The $\piz:\eta:\nbar$ ratio is fixed
 to the values obtained in the shower-shape analysis above. The measured
 signal contribution changes by $\SYSC$.
 \item 
 The resolution of the angular distance in the 
 $\rphi$ plane between the photon and the hadronic system is 
 approximately $5\%$. The systematic uncertainty is obtained 
 by changing the cut on this variable 
 by $\pm 5\%$ for the Monte Carlo events while leaving it 
 unchanged in the data. 
 This changes the cross-section by $\SYSANG$.
 \item 
 The contribution from photon-photon events with a photon from initial state
 radiation in the signal region was determined to be negligible using
 the BDK Monte Carlo~\cite{bib-bdk}.
 The contribution of events with final state radiation (FSR) to the
 data sample is estimated to be about $10-15\%$ using PHOJET. 
 The kinematic properties of these events are very similar to
 the single-resolved signal events. 
 Performing the measurement with and without taking into account
 the FSR contribution leads to a negligible change in the measured
 cross-section.
\end{itemize}

The resulting systematic uncertainties are summarised in Table~\ref{tab-sys}.

\section{Total cross-section}
The total inclusive cross-section for isolated prompt photon production 
with $\ptg > 3$~GeV and $|\etag|<1$ is obtained using
\begin{equation}
\sigma_{\rm tot}=\left(\frac{r}{\epsilon_{\rm single}} +
 \frac{1-r}{\epsilon_{\rm double}}\right)\frac{N_{\rm prompt}}{\lumi}. 
\end{equation}
The number of remaining prompt photons after the subtraction of
$\piz$, $\eta$, and $\nbar$ background 
is denoted by $N_{\rm prompt}$, and the single-resolved contribution 
$r$ is taken from the result of the fit to 
the two-dimensional $x_{\rm LL}^{\pm}$ distribution.
The efficiencies $\epsilon_{\rm single}$ and $\epsilon_{\rm double}$ 
are defined as
the total number of selected events divided by the total number of generated
events with an isolated prompt photon in the range 
$\ptg > 3.0\rm ~GeV$ and $|\etag|<1$ for the single
and double-resolved PYTHIA Monte Carlo samples. 
The efficiencies are determined to be $\epsilon_{\rm single}=\EPSSR$ and 
$\epsilon_{\rm double}=\EPSDR$.
This yields the cross-section
\begin{equation}
\XSECTION
\end{equation}
in the kinematic range defined by the anti-tagging condition.

\section{Differential cross-section}
To correct the measured differential cross-sections 
for acceptance losses and resolution effects in the detector, 
correction factors are determined in each bin using
the Monte Carlo simulation. 
The variables are first corrected event by event
for the average offset of the measured value in each bin
with respect to the generated value.
The resulting distribution is then multiplied by a bin-by-bin efficiency.
The dependence of the corrected distributions on the shape
of the generated Monte Carlo distributions is studied by reweighting the
generated distributions. The changes are found to be small compared
to the total uncertainties.

The inclusive differential cross-sections $\dsdpt$ and $\dsdeta$ 
for isolated prompt photon production are 
given in Table~\ref{tab-dxtg}.
The bin size is chosen to be significantly larger than the experimental 
resolution, which is about 200-300~MeV for $\ptg$. 
The $\etag$ resolution is much smaller than the
bin size due to the high ECAL granularity. 

The fractions of single and double-resolved
Monte Carlo events are determined using the measured 
$x_{\rm LL}^{\pm}$ distribution (Section~\ref{sec-sep}).
The systematic uncertainties related to the determination
of this ratio, to the ECAL calibration, and
to the modeling of the parton densities have been determined
for each bin separately, whereas all other systematic uncertainties are
added globally. 

\begin{figure}[htbp]
\vspace*{-2.cm}
   \begin{center}
      \mbox{
          \epsfxsize=15.0cm
          \epsffile{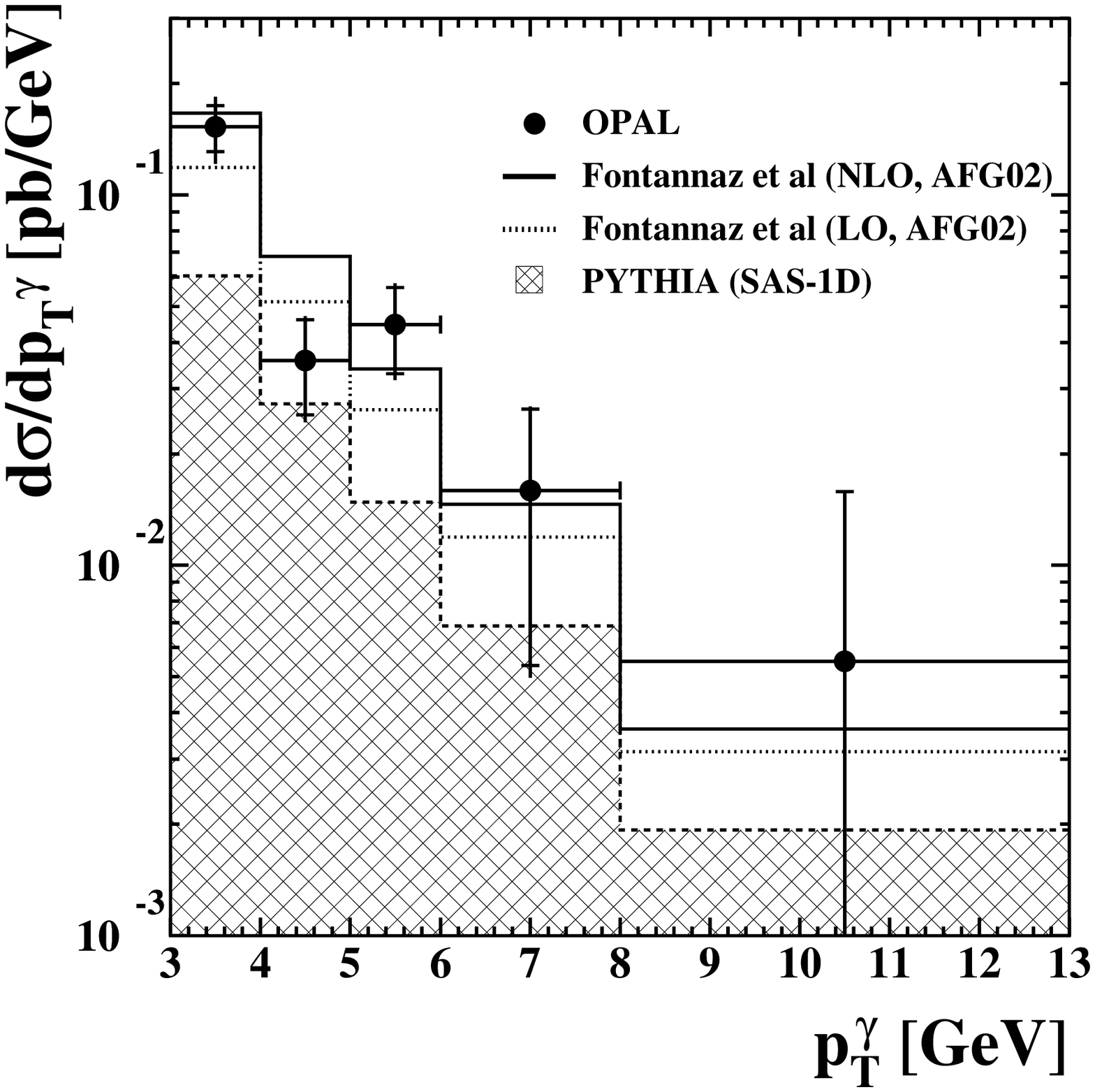}
           }
   \end{center}
\caption{
  Differential cross-section $\dsdpt$ for inclusive prompt photon production 
  in the kinematic range $|\etag|<1$ and $\ptg>3.0$~GeV. 
  The points represent data. 
  The inner error bars show the statistical uncertainty 
  and the outer error bars
  the total uncertainty. 
 }
\label{fig-diffp}
\end{figure}

\begin{figure}[htbp]
\vspace*{-2.cm}
   \begin{center}
      \mbox{
          \epsfxsize=15.0cm
          \epsffile{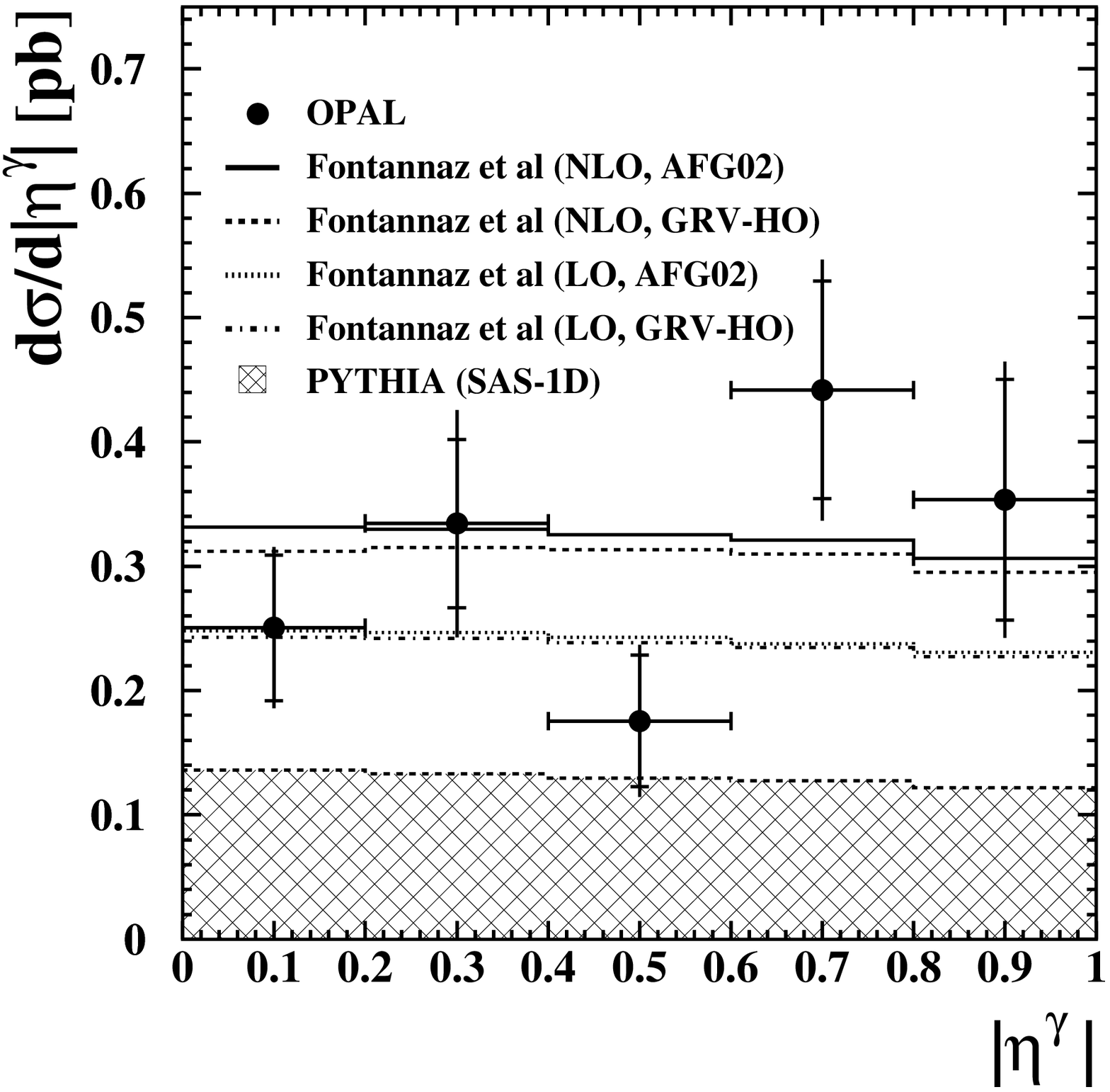}
           }
   \end{center}
\caption{ 
  Differential cross-section $\dsdeta$ for inclusive prompt photon production 
  in the kinematic range $|\etag|<1$ and $\ptg>3.0$~GeV. 
  The points represent data. 
  The inner error bars show the statistical uncertainty 
  and the outer error bars the total uncertainty. 
   }
\label{fig-diffe}
\end{figure}

In Figs.~\ref{fig-diffp} and~\ref{fig-diffe}, 
the inclusive differential cross-sections $\dsdpt$ and $\dsdeta$ 
are compared to the prediction of the Monte Carlo generator PYTHIA and
to the NLO calculation. 
For the PYTHIA simulation the SAS-1D parametrisation~\cite{bib-sas} 
is used with the original ratio of the single to double-resolved contribution 
given by PYTHIA. 
The NLO calculation uses the AFG02~\cite{bib-afg} 
parametrisation of the parton distributions of the photon
with $\Lambda^{(4)}_{\overline{\rm MS}}=300$~MeV and 
$\qmax=10$~GeV$^2$. 
The factorisation and renormalisation scales are set equal
to $\ptg$. 
The  calculated cross-sections are integrals over
the bins used in the analysis.

In both cases PYTHIA reproduces the shape of the distributions well but
underestimates the cross-sections, whereas
the NLO calculation~\cite{bib-fgh} describes well the shape and 
normalisation of the data. 
The differential cross-section $\dsdeta$ 
is independent of $|\etag|$ within the experimental uncertainties.
This effect is mainly due to the event-by-event
variation of the Lorentz boost from the $\gg$ system to the laboratory
system.
The NLO and LO calculations using
the AFG02~\cite{bib-afg} and GRV-HO~\cite{bib-grv} parametrisations
are also shown in Fig.~\ref{fig-diffe}. 
For this LO calculation the Born terms are used for the subprocess cross
sections together with the NLO strong coupling constant,
$\alpha_{\rm s}$, and the NLO parton distribution functions. 

The difference between the 
cross-sections using the two parametrisations is small
in comparison to the uncertainty of the data.

\begin{figure}[htbp]
   \begin{center}
      \mbox{
          \epsfxsize=15.0cm
          \epsffile{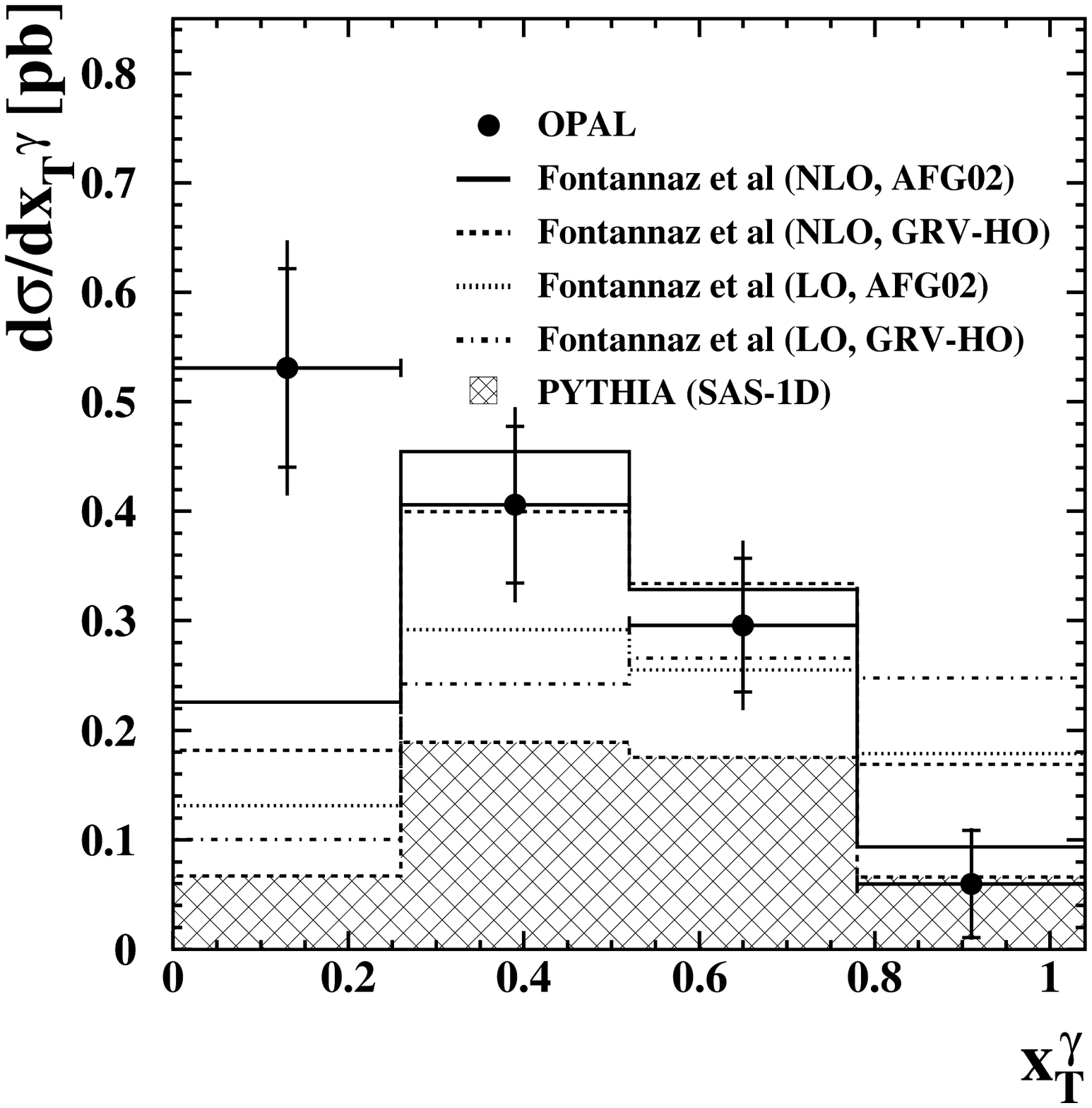}
           }
   \end{center}
\caption{ 
  Differential cross-section $\dsdxtg$ for inclusive prompt photon production 
  in the kinematic range $|\etag|<1$ and $\ptg>3.0$~GeV. 
  The points represent the data. 
  The inner error bars show the statistical uncertainty 
  and the outer error bars the total uncertainty. 
   }
\label{fig-diffxtg}
\end{figure}

The differential cross-section $\dsdxtg$ 
is shown in Fig.~\ref{fig-diffxtg} and the values are given 
in Table~\ref{tab-dxtg}.
The experimental resolution is in the range 0.05 to 0.15.
At low values of $\xtg$ the data lie about two standard deviations
above the NLO calculation which indicates a higher double-resolved 
contribution. 
The lowest kinematically accessible value is approximately 
$\xtg=2\ptg/W\simeq 0.1$.

\begin{figure}[htbp]
   \begin{center}
      \mbox{
          \epsfxsize=15.0cm
          \epsffile{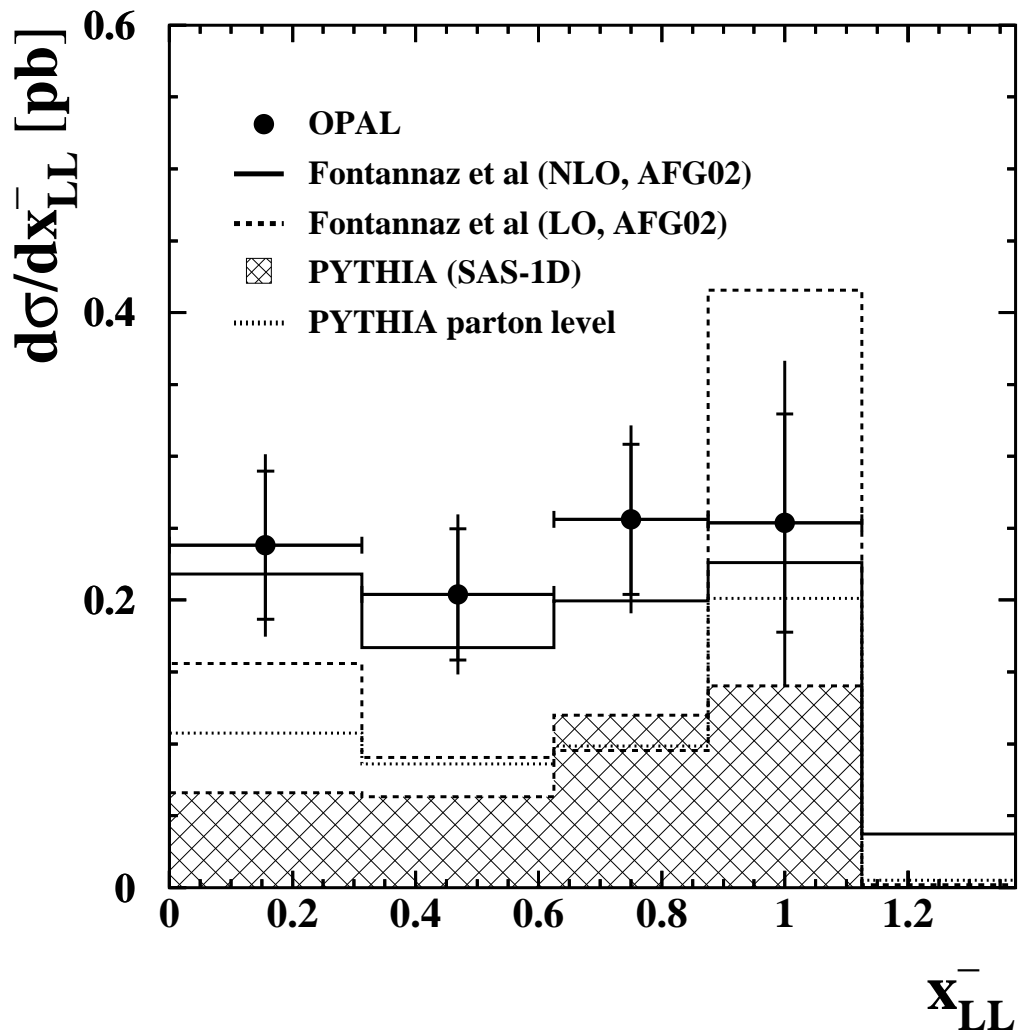}
           }
   \end{center}
\caption{ 
  Differential cross-section $\dsdxll$ for the production of a prompt 
  photon in association with at least one jet in the kinematic range 
  $|\etag|<1$, $\ptg>3.0$~GeV, $|\etajet|<2$, and $\ptjet>2.5$~GeV. 
  The points represent the data. 
  The inner error bars show the statistical uncertainty 
  and the outer error bars the total uncertainty. 
   }
\label{fig-diffxll}
\end{figure}

The differential cross-section $\dsdxll$ for the production of a prompt 
photon in association with at least one jet in the kinematic range 
$|\etag|<1$, $\ptg>3.0$~GeV, $|\etajet|<2$, and $\ptjet>2.5$~GeV
is shown in Fig.~\ref{fig-diffxll}. The values are given in 
Table~\ref{tab-dxll}.
If there are two jets associated with the photon, the photon plus 
jets cross-section 
is defined by keeping the jet with the highest $\ptjet$. 
The same procedure is used in the NLO calculation.

For the measurement of $\dsdxll$ an additional 
normalisation uncertainty of $10\%$ 
due to the jet requirement needs to be taken into account~\cite{bib-opaljet}. 
The main additional sources of this uncertainty are the energy scale
of the calorimeter, which is known to about $3\%$, and
the model dependence of the jet fragmentation~\cite{bib-opaljet}.

All measured cross-sections are given at the hadron level, i.e.,
no hadronisation corrections are applied to the data.
The hadron level cross-sections are compared to the parton level 
calculations by Fontannaz et al.~\cite{bib-fgh} in Fig.~\ref{fig-diffxll}.
A comparison of the PYTHIA parton and hadron cross-sections in 
Fig.~\ref{fig-diffxll} shows the size of hadronisation corrections
for this particular Monte Carlo model.

In the region $x_{\rm LL}^{\pm}>0.625$,
the peak around $x_{\rm LL}^{\pm}=1$ in the LO calculations is smeared
out towards lower values of $x_{\rm LL}^{\pm}$ both by the
hadronisation effects in the PYTHIA simulation and by
the higher order effects included in the NLO calculation. In this region
PYTHIA and the NLO calculation give a good description of
the data within the large uncertainties.
For $x_{\rm LL}^{\pm}<0.625$ hadronisation effects are expected to be
smaller. The NLO cross-section is larger than the LO cross-sections
in this kinematic range, in better agreement with the data.

\section{Conclusion}
The inclusive cross-section for the production of isolated prompt photons
in anti-tagged $\gg$ collisions 
is measured using the OPAL detector at LEP. Data with an integrated 
luminosity of $648.6 \rm ~pb^{-1}$ 
with centre-of-mass energies $\see$ from 183 GeV to 209 GeV are used.

The prompt photons are selected by requiring the isolation 
criterion of~\cite{bib-frixione}.
The signal and the background from $\piz,\eta$ and $\nbar$ production are
separated by a cluster shape analysis. 
In the kinematic region
$\ptg > 3.0 \rm ~GeV$ and $|\etag| < 1$, a total of $\NPROMPT$ events 
remain after background subtraction.
The total cross-section for inclusive isolated prompt photon production in  
the kinematic range defined by these cuts and by the
anti-tagging condition is measured to be 
\begin{equation}\XSECTION.\end{equation}
The anti-tagging cuts corresponds to $\qmax \approx 10$~GeV$^2$. 

Single and double-resolved events are separated using photon plus jet events, 
where the jets have been reconstructed with a cone jet algorithm. 
For the first time the differential cross-sections as a function of 
the transverse momentum, $\dsdpt$, the pseudorapidity, $\dsdeta$,
and the scaled transverse momentum, $\dsdxtg$, 
are measured and compared to the predictions of PYTHIA and to
a NLO calculation. In addition, we measure the differential
cross-section $\dsdxll$ for the production of a prompt 
photon in association with at least one jet in the kinematic range 
$|\etajet|<2$ and $\ptjet>2.5$~GeV.
The NLO calculation gives a better description
of the data than the LO calculation and the PYTHIA Monte Carlo,
especially at low $x_{\rm LL}^{\pm}$. 

\section*{Acknowledgements}

We thank Michel Fontannaz and Gudrun Heinrich for 
very helpful discussions and for providing the NLO calculations.
\par
We particularly wish to thank the SL Division for the efficient operation
of the LEP accelerator at all energies
 and for their close cooperation with
our experimental group.  In addition to the support staff at our own
institutions we are pleased to acknowledge the  \\
Department of Energy, USA, \\
National Science Foundation, USA, \\
Particle Physics and Astronomy Research Council, UK, \\
Natural Sciences and Engineering Research Council, Canada, \\
Israel Science Foundation, administered by the Israel
Academy of Science and Humanities, \\
Benoziyo Center for High Energy Physics,\\
Japanese Ministry of Education, Culture, Sports, Science and
Technology (MEXT) and a grant under the MEXT International
Science Research Program,\\
Japanese Society for the Promotion of Science (JSPS),\\
German Israeli Bi-national Science Foundation (GIF), \\
Bundesministerium f\"ur Bildung und Forschung, Germany, \\
National Research Council of Canada, \\
Hungarian Foundation for Scientific Research, OTKA T-038240, 
and T-042864,\\
The NWO/NATO Fund for Scientific Reasearch, the Netherlands.\\


\cleardoublepage
\begin{table}[htbp]
 \begin{center}
 \begin{tabular}{|l|c|}
  \hline
    source & uncertainty \\
  \hline    
      $\pi^0$ background                  & $\pz\SYSPI$ \\
      ECAL calibration                    & $\pz\SYSECAL$ \\
      ratio single to double-resolved contribution & $\pz\SYSSTD$\\
      parton density functions            & $\pz\SYSPDF$ \\
      HERWIG instead of PYTHIA            & $\pz\SYSHERWIG$ \\
      fixed $\eta : \nbar$ ratio          & $\pz\SYSETANBAR$\\
      re-weighting of single particle MC  & $\pz\SYSREW$\\
      minimum number of lead-glass blocks & $\pz\SYSMINBL$\\
      Monte Carlo statistics              & $\pz\SYSMCSTAT$\\
      binning effects                     & $\pz\SYSBIN$\\
      using $C$ parameter                 & $\pz\SYSC$\\
      resolution of $\phi^{\rm hadrons}-\phi^{\gamma}$ &  $\pz\SYSANG$\\ 
  \hline
      total systematic uncertainty     & $13.5\%$ \\
  \hline
  \end{tabular}
  \caption{Relative systematic uncertainties on the 
   total prompt photon cross-section, $\sigma_{\rm tot}$.}
  \label{tab-sys}
 \end{center}
\end{table}

\begin{table}[htpb]
 \begin{center}
 \begin{tabular}{|c|c|}
 \hline
 $\ptg$ [GeV]  &  $\dsdpt$ [pb/GeV] \\ \hline
 3.0 --  $\pz$4.0 &  $0.153 \pm  0.022 \pm  0.023$\\
 4.0 --  $\pz$5.0 &  $0.036 \pm  0.010 \pm  0.005$\\
 5.0 --  $\pz$6.0 &  $0.045 \pm  0.012 \pm  0.006$\\
 6.0 --  $\pz$8.0 &  $0.016 \pm  0.011 \pm  0.003$\\
 8.0 -- 13.0      &  $0.006 \pm  0.010 \pm  0.001$\\
 \hline
 \hline
   $|\etag|$  &   $\dsdeta$ [pb]\\ \hline
 0.0 --  0.2 &  $0.25 \pm  0.06 \pm  0.03$\\
 0.2 --  0.4 &  $0.33 \pm  0.07 \pm  0.06$\\
 0.4 --  0.6 &  $0.18 \pm  0.05 \pm  0.03$\\ 
 0.6 --  0.8 &  $0.44 \pm  0.09 \pm  0.06$\\
 0.8 --  1.0 &  $0.35 \pm  0.10 \pm  0.05$\\
 \hline
 \hline
   $\xtg$   &   $\dsdxtg$ [pb]\\ \hline
 0.00 --  0.26 &  0.53 $\pm$  0.09 $\pm$  0.07 \\
 0.26 --  0.52 &  0.41 $\pm$  0.07 $\pm$  0.05 \\
 0.52 --  0.78 &  0.30 $\pm$  0.06 $\pm$  0.05 \\
 0.78 --  1.04 &  0.06 $\pm$  0.05 $\pm$  0.01 \\
 \hline
 \end{tabular}
 \caption{Differential cross-sections $\dsdpt,\dsdeta$ and 
  $\dsdxtg$ for 
  $|\etag|<1$ and $\ptg>3$~GeV with the statistical and 
  systematic uncertainties.}
 \label{tab-dxtg}
 \end{center}
\end{table}
\begin{table}[hbpt]
 \begin{center}
 \begin{tabular}{|c|c|}
 \hline
   $x_{\rm LL}^-$  &   $\dsdxll$ [pb]\\ \hline
 0.0000 --  0.3125 &  0.24 $\pm$  0.05  $\pm$  0.04\\
 0.3125 --  0.6250 &  0.20 $\pm$  0.05  $\pm$  0.03\\
 0.6250 --  0.8750 &  0.26 $\pm$  0.05  $\pm$  0.04\\
 0.8750 --  1.1250 &  0.25 $\pm$  0.08  $\pm$  0.08\\
 \hline
 \end{tabular}
 \caption{Differential cross-section $\dsdxll$ for
  $|\etag|<1$, $\ptg>3.0$~GeV, $|\etajet|<2$, and $\ptjet>2.5$~GeV.
  The statistical and systematic uncertainties are also given.}
 \label{tab-dxll}
 \end{center}
\end{table}

\cleardoublepage


\end{document}